\documentstyle[aps]{revtex}

\begin{document}
\title{Macroscopical matter derived from Quantum Field Theory}
\author{G. Domenech\thanks{%
Fellow of the Consejo Nacional de Investigaciones Cientificas y Tecnicas
(CONICET) - levinas@iafe.uba.ar}}
\author{and M. L. Levinas\thanks{%
Fellow of the Consejo Nacional de Investigaciones Cientificas y Tecnicas
(CONICET) - levinas@iafe.uba.ar}}
\address{Instituto de Astronom\'\i a y F\'\i sica del Espacio (IAFE)\\
Casilla de Correo 67, Sucursal 28, 1428 Buenos Aires,\\
Argentina}
\maketitle

\begin{abstract}
We develop the WKB expansion to relate Quantum Field Theory variables with
those describing macroscopical matter. We find that, up to the first quantum
correction, free scalar fields correspond to perfect fluids with pressure.
We also find the law of motion for the associated particles which takes into
account deviations from the classical trajectories, showing that they do not
follow unaccelerated straight line trajectories.

\bigskip\ 


\noindent PACS number (s): 11.10.-z, 03.40.Gc
\end{abstract}

\section{Introduction}

The aim of this article is to find the macroscopical expression of the
energy-momentum tensor of a free scalar field derived from Quantum Field
Theory (QFT) up to the first quantum order in the WKB expansion, and to
study its macroscopical behavior. In previous works we established the
relation among QFT variables -amplitudes and phases- with macroscopic fluid
variables -proper energy and four-velocity- but considering only the lowest
order in $\hbar $ in the scheme of general curved space-times, in Riemannian
and non-Riemannian geometries. In both cases, free scalar matter corresponds
to perfect fluids without pressure following geodesics[1, 2]. In fact, this
statement also holds for flat Minkowski space-time where geodesics must be
understood as straight non-accelerated trajectories (inertial principle). In
the present article we prove that up to the following order, free scalar
fields correspond to perfect fluids {\it with pressure}. We also find the
corresponding law of motion which shows dispersions from inertial motion. Up
to our knowledge, these are new results which give a deeper insight in the
relation between second quantization and standard description of
macroscopical matter. We believe that these results contribute to place QFT
as the ultimate description of matter structure.

The paper is organized as follows: in section II we expand the
energy-momentum tensor of the free scalar field. In section III, we find the
relation between microscopical and macroscopical variables and prove that
free scalar matter must be represented by a perfect fluid with pressure when
quantum corrections are taken into account. In IV we find the equation which
relates the WKB amplitudes that are involved in the corresponding
definitions of proper energy densities and pressures; this section is also
devoted to find the first quantum correction to the classical equations of
motion. Finally, in V, we make a summary of our results.

\section{The WKB expansion of the energy-momentum tensor}

In order to study the relationship between quantum and classical dynamics,
we start considering the complex wave function $\Psi $ in terms of the path
integral formalism. In the first quantization scheme, the wave function of
the particle may be expressed as:

\begin{equation}
\Psi (x_o,x_f)=N\int e^{iS_{cl}[x]/\hbar }{\cal D}x  \label{1}
\end{equation}

\noindent where $S_{cl}[x]$ is the classical action which is a functional of
the trajectory $x(t)$ depending on the initial and final fixed data ($%
x_o,x_f $) and $N$ is a normalization factor. If $S_{cl}$ is the action of a
non-relativistic particle, then $\Psi $ corresponds to a wave function which
satisfies the Schroedinger equation [3]. On the other hand, when we work in
the scheme of second quantization, $\Psi $ must be taken to be a field
satisfying the relativistic field equation obtained for the corresponding
representation of the Lorentz group. Field equations arise minimizing action 
$S[\Psi ]$ with respect to the field:

\begin{equation}
\frac{\delta S[\Psi ]}{\delta \Psi }=0  \label{2}
\end{equation}

The expansion of eq. (1) in powers of $\hbar $ reads:

\begin{equation}
\Psi =Ne^{iS_{cl}[x_{cl}]/\hbar }\int e^{\int \frac 12\delta
S_{cl}[x]/\delta x_1\delta x_2{\cal D}x_1{\cal D}x_2+...}\text{ }{\cal D}%
x=\sum_{n=0}(-i\hbar )^n\Psi _ne^{iS_{cl}/\hbar }  \label{3}
\end{equation}
with $x_{cl}$ the classical trajectory. This expression is known as the WKB
expansion. For any group representation, this expansion means that:

\begin{equation}
S[\Psi ]=S_{cl}[x]+{\cal O}(\hbar )  \label{4}
\end{equation}

\noindent So, the $\hbar =0$ limit of $S[\Psi ]$ in a first quantization
scheme leads to $S_{cl}[x],$ in the same way as the $\hbar =0$ limit of the
effective action conduces, in a second quantization scheme, to $S[\Psi ].$

We are interested in analyzing the case of a complex scalar field $\Phi .$
Its action reads:

\begin{equation}
S[\Psi ]=\int {\cal L}(\Phi ,\partial \Phi )\text{ }d^4x{\cal =}\int \frac 12%
(\partial ^\mu \Phi \partial _\mu \Phi ^{*}+\frac{m^2}{\hbar ^2}\Phi \Phi
^{*})\text{ }d^4x  \label{5}
\end{equation}

The corresponding energy-momentum tensor is: 
\begin{equation}
T^{\mu \nu }\equiv \frac{\partial {\cal L}}{\partial \partial _\mu \Phi }%
\partial ^\nu \Phi +\frac{\partial {\cal L}}{\partial \partial _\mu \Phi ^{*}%
}\partial ^\nu \Phi ^{*}-{\cal L}\eta ^{\mu \nu }  \label{6}
\end{equation}

\begin{quotation}
\noindent which conduces to: 
\begin{equation}
T^{\mu \nu }=\partial ^{(\mu }\Phi \partial ^{\nu )}\Phi ^{*}-\frac 12\eta
^{\mu \nu }\partial ^\lambda \Phi \partial _\lambda \Phi ^{*}-\frac 12\eta
^{\mu \nu }\frac{m^2}{\hbar ^2}\Phi \Phi ^{*}  \label{7}
\end{equation}

The WKB expansion (3) for the scalar field reads: 
\begin{equation}
\Phi =\sum_{n=0}(-i\hbar )^n\varphi _ne^{iS_{_{cl}}/\hbar }  \label{8}
\end{equation}

Now, using expansion (8) in expression (7), and after a straightforward
computation, we find that up to order $1/\hbar $, the energy momentum tensor
can be written as:
\end{quotation}

\begin{center}
$T^{\mu \nu }=\frac 1{\hbar ^2}\partial ^\mu S\partial ^\nu S\varphi
_o\varphi _o^{*}+\frac 1\hbar \{\partial ^{(\mu }S$ Im$(\varphi _o\partial
^{\nu )}\varphi _o^{*})-2\partial ^\mu S\partial ^\nu S$ Im($\varphi
_o\varphi _1^{*})+$

\begin{equation}
\eta ^{\mu \nu }[(\partial ^\lambda S\partial _\lambda S)\text{ Im}(\varphi
_o\varphi _1^{*})+(\partial ^\lambda S)\text{ Im}(\varphi _o\partial
_\lambda \varphi _o^{*})+m^2\text{ Im}(\varphi _o\varphi _1^{*})]\}
\label{9}
\end{equation}
\end{center}

\begin{quotation}
\noindent where $S_{cl}=S,$ and Im means ``imaginary part{\it ''}. It is
easy to verify that this symmetric quantity is conserved, i.e. satisfies: 
\begin{equation}
\partial _\mu T^{\mu \nu }=0  \label{10}
\end{equation}
\end{quotation}

\section{The Scalar Field as a Perfect Fluid}

\begin{quotation}
Let us analyze the $\hbar ^{-2}$ order (the first term) of expression (9).
We may write the corresponding term as

\begin{equation}
T^{(1/\hbar ^2)\mu \nu }=\hat \rho u^\mu u^\nu  \label{11}
\end{equation}

\noindent where

\begin{equation}
\hat \rho =\frac{m^2}{\hbar ^2}\varphi _o\varphi _o^{*}  \label{12}
\end{equation}

\noindent is the proper energy and $u^\mu $ is the timelike vector
orthogonal to the hypersurface of constant phase $S$, where $\partial _\mu
S=P^\mu $ and $S$ is understood as the Hamilton principal function. In the
free case, canonical momentum $P^\mu $ coincides with the ordinary momentum $%
p^\mu $, i.e.:

\begin{equation}
P^\mu =p^\mu =mu^\mu  \label{13}
\end{equation}

Four-velocity $u^\mu $ satisfies $u^\mu u_\mu =-1.$ Relation (11) shows that 
$T^{(1/\hbar ^2)\mu \nu }$ is nothing but the energy-momentum tensor of a
dust.

Using (13) in (9) we are able to write the energy-momentum tensor showing
explicitly the four-velocity components: 
\begin{eqnarray}
T^{\mu \nu } &=&\frac{m^2}{\hbar ^2}\varphi _o\varphi _o^{*}u^\mu u^\nu -2%
\frac m\hbar u^{(\mu }\text{Im}(\varphi _o\partial ^{\nu )}\varphi _o^{*})
\label{14} \\
&&-2\frac{m^2}\hbar u^\mu u^\nu \text{ Im(}\varphi _o\varphi _1^{*})-\frac m%
\hbar u^\lambda \text{ Im}(\partial _\lambda \varphi _o\cdot \varphi
_o^{*})\eta ^{\mu \nu }  \nonumber
\end{eqnarray}

From (14) we can see that at a point where the observer is at rest (proper
observer), i.e.: $\tilde u^\mu =(-1,0,0,0)$, the components of $T^{\mu \nu
}, $ which we denote as $\tilde T^{\mu \nu },$ are:

\begin{equation}
\tilde T^{oo}=\frac{m^2}{\hbar ^2}\varphi _o\varphi _o^{*}-\frac m\hbar [2m%
\text{ Im}(\varphi _o\varphi _1^{*})+\text{Im(}\varphi _o\dot \varphi
_o^{*})]  \label{15}
\end{equation}

\begin{equation}
\tilde T^{oi}=0  \label{16}
\end{equation}

\begin{equation}
\tilde T^{ij}=\frac m\hbar \text{Im(}\varphi _o\dot \varphi _o^{*})\text{ }%
\delta ^{ij}  \label{17}
\end{equation}

\noindent (with $i$, $j$, $...$ $=$ $1$, $2$, $3$). This will be useful in
order to find the general expression in any Lorentzian frame.

Once we have written the components of $T^{\mu \nu },$ we see that it is
possible to define:

\begin{equation}
\rho =\frac{m^2}{\hbar ^2}\varphi _o\varphi _o^{*}-\frac m\hbar [2m\text{ Im}%
(\varphi _o\varphi _1^{*})+\text{Im(}\varphi _o\dot \varphi _o^{*})]\equiv 
\hat \rho -\breve \rho  \label{18}
\end{equation}

\begin{equation}
p=\frac m\hbar \text{Im(}\varphi _o\dot \varphi _o^{*})  \label{19}
\end{equation}

\noindent which are the proper energy density and the pressure,
respectively. Here $\cdot $ $\equiv d/d\tau $, $\tau $ being the proper
time, $\hat \rho $ is given by (12) while $\breve \rho $ is $\breve \rho =%
\frac m\hbar [2m$ Im$(\varphi _o\varphi _1^{*})+$Im($\varphi _o\dot \varphi
_o^{*})].$ So, the components of $\tilde T^{\mu \nu }$ read:

\begin{equation}
\tilde T^{oo}=\rho  \label{20}
\end{equation}

\begin{equation}
\tilde T^{oi}=0  \label{21}
\end{equation}

\begin{equation}
\tilde T^{ij}=p\text{ }\delta ^{ij}  \label{22}
\end{equation}

Expressions (20-22) represent the fluid that a proper observer would
describe as isotropic, i.e. a perfect fluid.

Now we want to generalize expressions (20-22) from this system to any
Lorentz frame. To do this we analyze how $T^{\mu \nu }$ may be `measured '
in a reference frame at rest in the laboratory supposing that the fluid
appears to be moving at each space-time point with velocity $v^i$. This
velocity is related to the spatial components of the four velocity $u^i$ of
the fluid via:

\[
v^i=\gamma ^{-1}\cdot u^i\text{ } 
\]
where $\gamma =(1-\bar v^2/c^2)^{-1/2}$ is the boost that transforms Lorentz
reference frames among them. Applying this boost to components (20-22) of $%
T^{\mu \nu },$ we obtain:

\begin{equation}
T^{oo}=(\rho +p\bar v^2)\gamma ^2  \label{23}
\end{equation}

\begin{equation}
T^{oi}=(\rho +p)v^i\gamma ^2  \label{24}
\end{equation}

\begin{equation}
T^{ij}=p\delta ^{ij}+(\rho +p)v^iv^j\gamma ^2  \label{25}
\end{equation}

\noindent with $\rho $ and $p$ given by (18) and (19), respectively.

Combining (23), (24) and (25), we see that in any Lorentz frame, expression
(14) reads:

\begin{equation}
T^{\mu \nu }=\rho u^\mu u^\nu +p(\eta ^{\mu \nu }+u^\mu u^\nu )  \label{26}
\end{equation}

\noindent i.e.: a perfect fluid. In conclusion: expression (9) for $T^{\mu
\nu }$\ which is given by QFT, corresponds to a perfect fluid that can be
written only in terms of macroscopical variables.

In order to connect proper energy density and pressure we would only need an
equation of state. The structure of this equation depends on the properties
of the system. In the particular case in which temperature is an independent
variable, pressure is a function of $\rho _o$ and $\Pi $:

\begin{equation}
p=p(\rho _o,\Pi )  \label{27}
\end{equation}

where 
\begin{equation}
\rho =\rho _o(1+\Pi )  \label{28}
\end{equation}

$\rho _o$ being the proper mass density and the product $\rho _{o^{.}}\Pi $,
the internal energy density.
\end{quotation}

\section{Relation between the WKB amplitudes and equations of motion}

In QFT, field equations are obtained minimizing the effective action with
respect to the field (equation (2)). For the scalar representation $\Phi $,
this conduces to the Klein-Gordon equation:

\begin{center}
\begin{equation}
\Box \Phi +\frac{m^2}{\hbar ^2}\Phi =0  \label{29}
\end{equation}
\end{center}

\noindent which in terms of the WKB expansion of the field, reads:

\begin{equation}
\sum_n[\hbar ^2\Box \phi _n+i\hbar (2\partial ^\mu \phi _n\partial _\mu
S+\phi _n\Box S)-\phi _n\partial ^\mu S\partial _\mu S)(-i\hbar
)^n]=-\sum_nm^2\phi _n(-i\hbar )^n  \label{30}
\end{equation}

\begin{quotation}
Assuming that (30) must be satisfied order by order in $\hbar ,$ we obtain
the differential equations which are needed to describe the perfect fluid
via eqs. (18) and (19):

\begin{equation}
n=0\hspace{1.0in}\partial ^\mu S\partial _\mu S+m^2=0  \label{31}
\end{equation}

\begin{equation}
n=1\hspace{1.0in}2i\partial ^\mu \phi _o\partial _\mu S+i\phi _o\Box S+\phi
_1(\Box S+m^2)=0  \label{32}
\end{equation}

Equation (31) provides no new information: it is trivially satisfied because 
$\partial ^\mu S=P^\mu =mu^\mu $ and $u^\mu u_\mu =-1$. Equation (32) may be
written as:

\begin{equation}
2u^\mu \partial _\mu \phi _o+(\phi _o-i\phi _1)\partial _\mu u^\mu -im\phi
_1=0  \label{33}
\end{equation}

\noindent and relates the first two amplitudes of the WKB\ expansion of the
field.
\end{quotation}

In order to find the equations of motion corresponding to scalar particles,
we compute equation (10) using (19) and obtain:

\begin{equation}
\partial ^\mu P+\partial _\nu [(\rho +P)u^\mu u^\nu ]=0  \label{34}
\end{equation}

Keeping only those equations corresponding to the spatial components of the
velocity $\bar v,$ we find the following vectorial equation:

\begin{equation}
(\rho +P)\cdot [\frac{\partial \bar v}{\partial t}+(\bar v\cdot \bar \nabla )%
\bar v]=-\gamma ^{-2}[\bar \nabla P+\bar v\frac{\partial P}{\partial t}]
\label{35}
\end{equation}

\noindent where we have used the fact that $u^i=v^i\cdot u^o$. In eq. (35),
different orders in $1/\hbar ^n$ are involved due to the corresponding
expressions of $\rho $ and $P$ (see (18) and (19)). So, for the first two
orders, we obtain two equalities. Up to the $1/\hbar ^2$ (main) contribution
we have :

\begin{equation}
\frac{\partial \bar v}{\partial t}+(\bar v\cdot \bar \nabla )\bar v=0
\label{36}
\end{equation}

\noindent  Taking into account the $1/\hbar $ correction, eq. (36) becomes:

\begin{equation}
(\hat \rho -\breve \rho +P)\cdot [\frac{\partial \bar v}{\partial t}+(\bar v%
\cdot \bar \nabla )\bar v]=-\gamma ^{-2}[\bar \nabla P+\bar v\frac{\partial P%
}{\partial t}]  \label{37}
\end{equation}

The first equation tells us that, at the highest order in the WKB expansion,
the free massive scalar field is represented by particles with null
acceleration, i.e., their trajectories are straight lines. This is the
classical limit of the approach and represents the inertial principle. But
we can see from eq. (37) that quantum corrections to the straight line
appear at the next order. That is to say: free matter follows trajectories
that deviate from straight lines and from non-accelerated uniform motion.

We want to recall that equation (37) may be considered the relativistic
second law of Newton: $F^\mu =m\cdot du^\mu /d\tau $, where quantum
corrections to $\rho $ and the appearance of $P$ (expressions (18) and (19)
respectively), are responsible for deviations from straight lines with no
need of external forces nor of quantum potentials (as, for example, in the
Bohm's scheme; see[4]). Equation (37) reveals the nature of the internal
structure of matter and how it acts in order to produce quantum effects.

\section{Summary}

We have developed the WKB expansion of the scalar field up to the first
quantum correction and studied the relations between QFT\ description of
matter and macroscopic fluids. Our aim was to find the behaviour of this
matter in a macroscopical picture, asumming quantum corrections. Our main
results are the following:

\noindent  - at the first quantum approximation, free scalar fields
correspond to perfect fluids with pressure,

\noindent  - the corresponding law of motion implies deviations from
straight unaccelerated motion.

\begin{quotation}
\bigskip\ 
\end{quotation}

\section{References}

\noindent [1] M. Castagnino, G. Domenech, M. Levinas and N. Um\'erez,
Class.Quant.Grav. {\bf 6} (1989) L 1.

\noindent [2] G. Domenech, M. Levinas and N. Um\'erez, Phys. Lett.{\bf \
137A }(1989) 17.

\noindent [3] R. Feynman and A. Hibbs, {\it Quantum Mechanics and Path
Integral}, New York, Mc.Graw-Hill, 1965, chap. 4.

\noindent [4] D. Bohm, B. Hiley and P. Kaloyerou, Phys. Rep.144 No. 6 (1987)
321.

\end{document}